\documentclass[journal=aelccp, manuscript=letter]{achemso}

\usepackage{amsmath}
\usepackage{hyperref}
\usepackage{cleveref}
\usepackage{comment}

\newcommand*\kB{k_{\text{B}}} 
\newcommand*\Ei{E_{\text{i}}} 
\renewcommand*\ni{n_{\text{i}}} 
\newcommand*\Et{E_\text{t}} 
\newcommand*\vthn{v_{\text{th,}n}} 
\newcommand*\vthp{v_{\text{th,}p}} 
\newcommand*\tauSRH{\tau_{\text{SRH}}} 
\newcommand*\Ea{E_{\text{a}}} 
\newcommand*\Ean{E_{\text{a,n}}} 
\newcommand*\Eap{E_{\text{a,p}}} 

\author{Rachel C. Kurchin}
\affiliation{Massachusetts Institute of Technology, Cambridge, MA, USA}
\alsoaffiliation{now at Carnegie Mellon University, Pittsburgh, PA, USA}
\author{Jeremy R. Poindexter}
\altaffiliation{now at MiaSol\'e Hi-Tech Corp., Santa Clara, California, USA}
\affiliation{Massachusetts Institute of Technology, Cambridge, MA, USA}
\author{Ville V\"ah\"anissi}
\author{Hele Savin}
\affiliation{Department of Electronics and Nanoengineering, Aalto University, Espoo, Finland}
\author{Carlos del Ca\~nizo}
\affiliation{Instituto de Energ\'ia Solar, Universidad Polit\'ecnica de Madrid, Madrid, Spain}
\author{Tonio Buonassisi}
\affiliation{Massachusetts Institute of Technology, Cambridge, MA, USA}
\email{buonassisi@mit.edu}

\title[Fe_Si_Bayes]
  {How much physics is in a current-voltage curve? Inferring defect properties from photovoltaic device measurements}

\begin{document}

\begin{tocentry}

\centering
\includegraphics[width=0.85\textwidth]{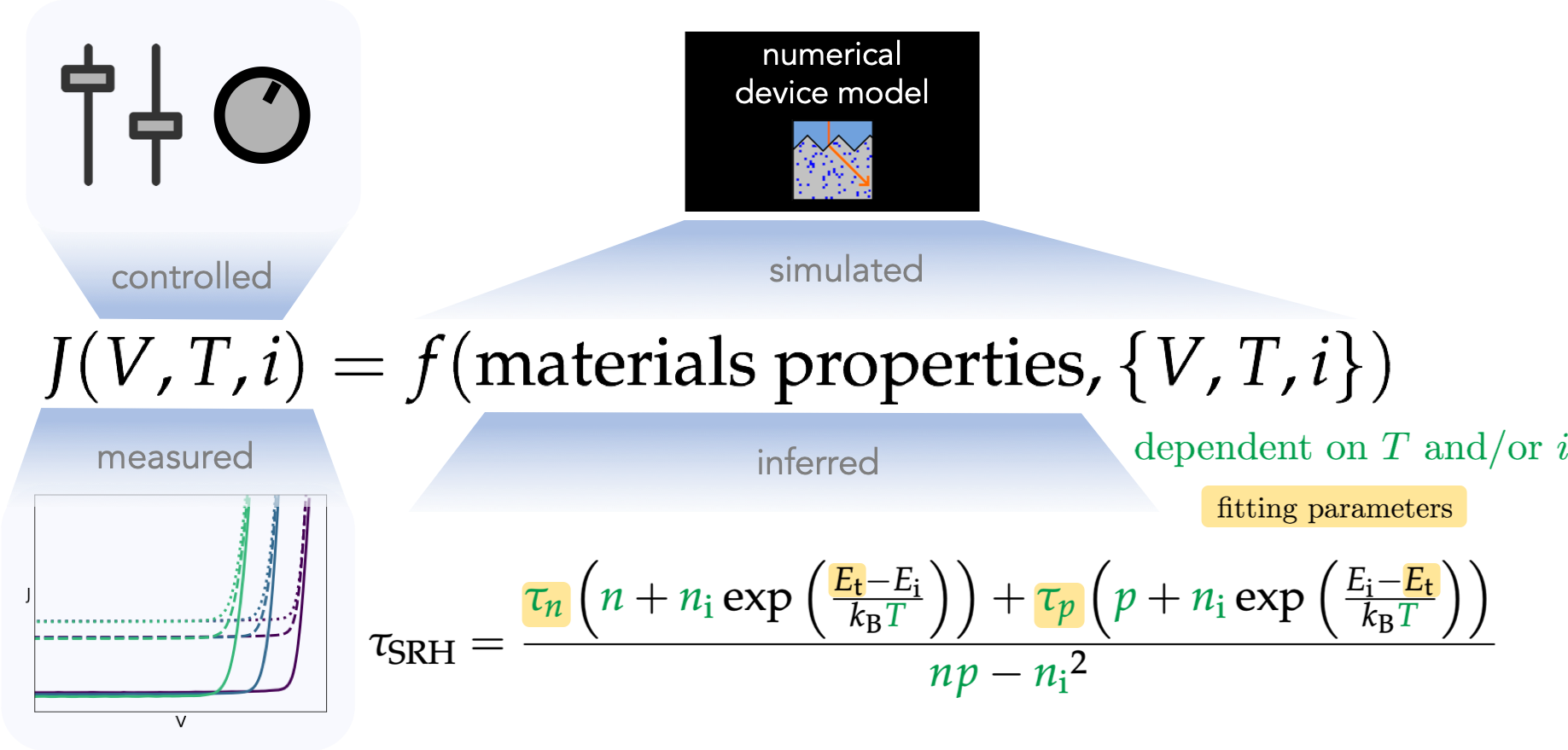}

\end{tocentry}

\begin{abstract}
    Defect-assisted recombination processes are critical to understand, as they frequently limit photovoltaic (PV) device performance. However, the physical parameters governing these processes can be extremely challenging to measure, requiring specialized techniques and sample preparation. And yet the fact that they limit performance as measured by current-voltage (JV) characterization indicates that they must have some detectable signal in that measurement. In this work, we use numerical device models that explicitly account for these parameters with high-throughput JV measurements and Bayesian inference to construct probability distributions over recombination parameters, showing the ability to recover values consistent with previously-reported literature measurements. The Bayesian approach enables easy incorporation of data and models from other sources; we demonstrate this with temperature dependence of carrier capture cross-sections. The ability to extract these fundamental physical parameters from standardized, automated measurements on completed devices is promising for both established industrial PV technologies and newer research-stage ones.
\end{abstract}

Recombination mediated by point defects is a performance-limiting mechanism in many photovoltaic (PV) technologies \cite{Schubert2015, Collord2015, Durose2019}. Identifying and characterizing these defects is essential to mitigating their effects. Typically, defect characterization is performed on wafers or semifabricates using temperature- and/or injection- dependent lifetime spectroscopy (TIDLS) \cite{lifetime_spec, Morishige2017-1}, deep level transient spectroscopy (DLTS) \cite{Lang1974, dlts, Wickramaratne2018}, and related spectroscopy techniques. However, these techniques are time-consuming, and the deep expertise necessary to master them is rare. Measurements on semifabricates may not be representative of finished devices, as final processing can affect defect populations. With the maturation of data-science methods, we explore the possibility of extracting defect information directly from non-destructive electrical device measurements.

Any defects detrimental to device performance should by definition have a signature in device performance such as current-voltage (JV) measurements. However, such a signal is convoluted with those from so many other physical processes that it cannot be extracted or interpreted through a simple fitting approach, as the fit would be underconstrained. However, by combining current-voltage measurements at a range of temperatures and light intensities (JVTi) with physics-based device models \cite{Haug2015, Haug2016, SCAPS} and Bayesian statistics, these signals can be disentangled, providing fits for many types of underlying parameters, often with greater precision than direct characterization allows. 

We previously demonstrated this approach to measure materials properties such as minority carrier mobility and lifetime in a finished SnS solar cell. \cite{SnSJoule} The Bayesian framework enables quantifying parameter-specific uncertainty as well as observing emergent relationships between parameters (such as mobility-lifetime product). In this work, we apply this approach to extract defect-assisted recombination parameters for interstitial iron in silicon, obtaining results consistent with reported literature values. Our results demonstrate a novel approach to extract defect properties from inexpensive measurements of completed devices, demonstrating promise for characterization of both established and novel PV technologies.

Defect-assisted recombination is described by the Shockley-Read-Hall (SRH)\cite{SR,H} equation, where the SRH lifetime $\tauSRH$ is given by:

\begin{equation}
    \tauSRH =\frac{\tau_n\left(n+\ni\exp\left(\frac{\Et-\Ei}{\kB T}\right)\right)+\tau_p\left(p+\ni\exp\left(\frac{\Ei-\Et}{\kB T}\right)\right)}{np-\ni^2},
    \label{SRH}
\end{equation}

\noindent where $n$, $p$ are the concentrations of electrons and holes, respectively, $\ni$ is the intrinsic electron concentration, $E_t$ is the energy level of the defect (trap), $\Ei$ is the intrinsic Fermi level, $T$ is temperature, $\kB$ is Boltzmann’s constant, and the lifetime parameters $\tau_n$, $\tau_p$ are given by:

\begin{equation}
    \tau_n = \frac 1{N_t\sigma_n\vthn}
    \label{taun}
\end{equation}
\begin{equation}
    \label{taup}
    \tau_p = \frac 1{N_t\sigma_p\vthp},
\end{equation}

\noindent where $N_t$ is the defect concentration, $\sigma_n$ and $\sigma_p$ are the defect capture cross sections for electrons and holes, respectively, and $\vthn$, $\vthp$ are the thermal velocities of electrons and holes, respectively. 

Interstitial iron is one of the most detrimental (and hence best characterized) point defects in silicon PV devices. In this work, we seek to characterize $\tau_n$, $\tau_p$, and $E_t$ from $JVTi$ measurements. Varying temperature and illumination intensity is critical to distinguish the influences of different defect parameters. These dependencies on experimental conditions are encoded in PC1D \cite{Haug2015, Haug2016}, the device simulation software we chose for this study. (For a visualization of the impact of various parameters, see SI \Cref{figS3}) In general, carrier concentrations depend linearly on light intensity. PC1D does not explicitly include temperature dependence of capture cross-sections; we account for this ourselves and the mathematical model is discussed below (see \Cref{sigma_n_Arr,sigma_p_Arr}).

\begin{figure}
    \centering
    \includegraphics[width=0.9\textwidth]{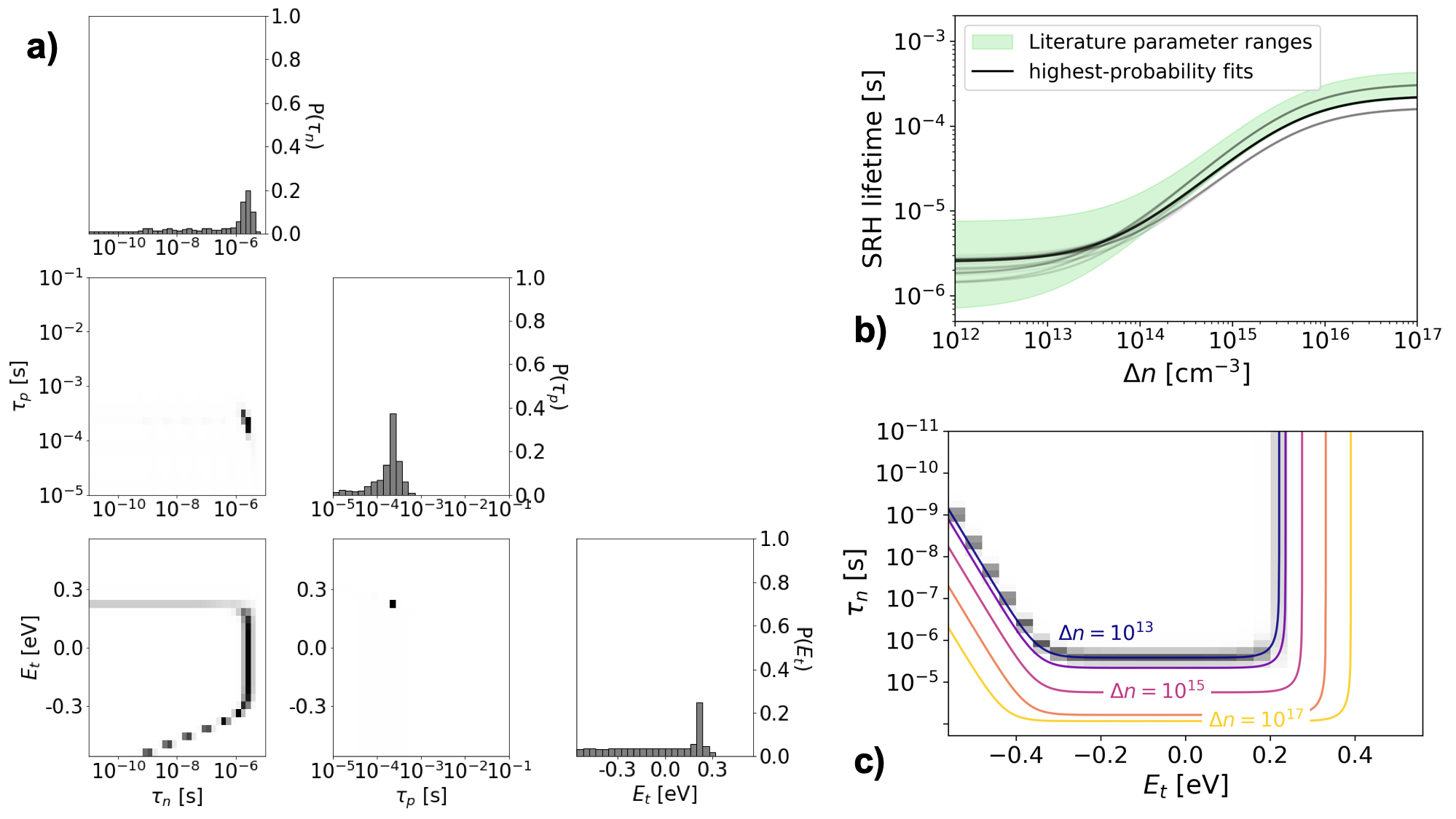}
    \caption{Visualizations of results of three-parameter fit at 300K. a) Probability distribution, with single-variable marginalizations along the diagonal and two-variable marginalizations off-diagonal. b) Simulated SRH lifetime vs. injection for the highest probability sets of parameters. Intensity of lines proportional to probability, top 80 parameter sets (corresponding to XX\% of total probability mass) shown. Green region shows simulated data based on ranges of parameters found in the literature. c) Marginalization between $E_t$ and $\tau_n$ from (a) with calculated iso-injection curves overlaid.}
    \label{fig1}
\end{figure}

Using $JV$ measurements taken from 175--300 K and 0.09--1 Sun, we first construct probability distributions over $\tau_n$, $\tau_p$, and $E_t$ at each temperature separately. An example (at 300 K) is plotted in \Cref{fig1}a. Next, we choose the highest-probability points in this three-dimensional parameter space and use them to construct simulated SRH lifetime curves as a function of carrier injection level, shown in \Cref{fig1}b. Also shown (in green) is the range corresponding to the ranges of parameters reported in the literature~\cite{istratov, lifetime_spec} and constructed using tabulated values for thermal velocities in silicon~\cite{Green1990} and previously-characterized defect densities on this sample~\cite{pdg_si}. The simulated curves from this study are well within the literature ranges.

\Cref{fig1}c shows the marginal distribution between $\tau_n$ and $E_t$ from \Cref{fig1}a, with iso-injection curves overlaid. These were constructed using a fixed $\tau_p$ value, a reasonable assumption given the highly concentrated probability distribution over this parameter seen in \Cref{fig1}a. $\tauSRH$ was fixed to the logarithmic average over the range computed from literature parameters in \Cref{fig1}b, and then \Cref{SRH} inverted to give a relationship between $\tau_n$ and $E_t$. The results are consistent with the fact that these devices should be in low injection under the illumination levels used. This analysis again demonstrates that similar information to lifetime spectroscopy can be gleaned from our approach.

\begin{figure}
    \centering
    \includegraphics[width=0.5\textwidth]{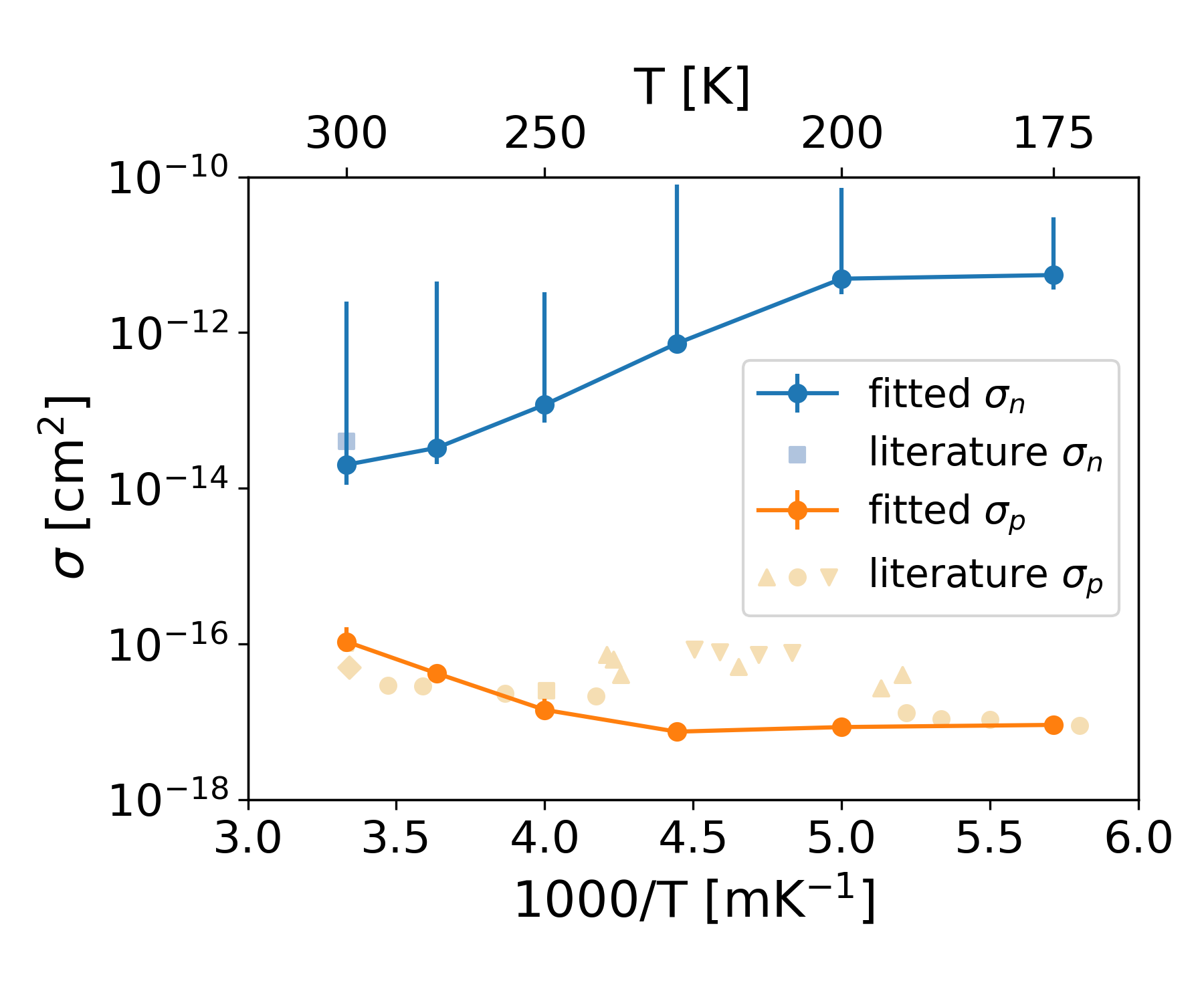}
    \caption{Inferred capture cross sections vs. temperature compared to literature values. Scatter points are from means of $\tau$ probability distribution, error bars are from interquartile ranges. Different symbols represent different sources of literature data.}
    \label{fig2}
\end{figure}

As alluded to above, because thermal velocities in silicon are tabulated and trap density in this sample has been characterized, we can directly extract capture cross sections (see \Cref{taun,taup}). Converting $\tau$’s to $\sigma$’s in this way and plotting against literature data from a variety of sources \cite{Lemke1981, Lemke1981-1, Wunstel1982, Brotherton1985, Indusekhar1986, Gao1991, Rein2005} and collated by Ref.~\citenum{istratov}, yields \Cref{fig2}, which shows capture cross sections for each carrier against temperature, with error bars computed as the interquartile range of the single-variable marginalization from the probability distribution at that temperature. Acquisition methods in literature data include DLTS, thermally stimulated capacitance (TSCAP, a predecessor technique to DLTS), and Hall effect.

A widely-accepted model for carrier capture is as a thermally activated process~\cite{Passler1978, istratov} Implementing such a model allows an Arrhenius relation to be used for each capture cross-section, introducing two new parameters for each carrier: a prefactor $\sigma_0$ and an activation energy $\Ea$:

\begin{equation}
    \sigma_n = \sigma_{n0}e^{\Ean/\kB T}
    \label{sigma_n_Arr}
\end{equation}
\begin{equation}
    \sigma_p = \sigma_{p0}e^{\Eap/\kB T}.
    \label{sigma_p_Arr}
\end{equation}

\begin{figure}
    \centering
    \includegraphics[width=0.9\textwidth]{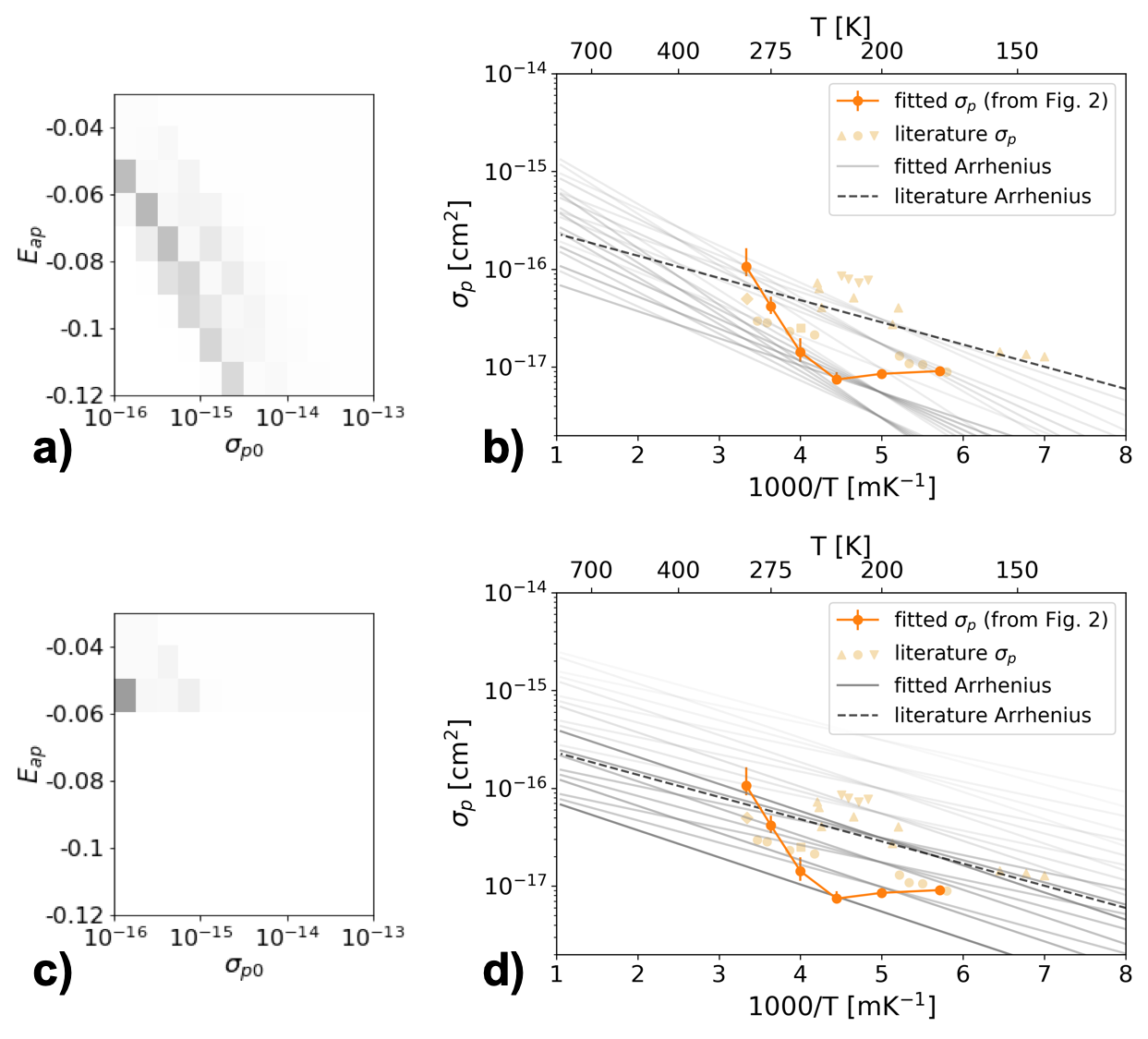}
    \caption{a) $\Eap$-$\sigma_{p0}$ marginalization from five-parameter fit. (Excerpt from \Cref{figS1}) b) $\sigma_p$ data from \Cref{fig2} with inferred Arrhenius fits, intensity of line proportional to probability of parameters, and Arrhenius fit from literature. c) Marginalization from (a) conditioned on $\Eap$ value being within .01 eV of the literature value of -0.045 eV. d) Same plot as (b) but for the marginalized PMF from (c). (top 20 Arrhenius fit parameter sets plotted in both (b) and (d))}
    \label{fig3}
\end{figure}

The parameter space is now five-dimensional, but we can also constrain a single posterior distribution using all the data rather than needing separate fits at each temperature. The probability distribution resulting from this analysis is shown in \Cref{figS1}. Moving forward, we focus on $\sigma_p$ in literature data comparisons, because significantly more data has been reported than for $\sigma_n$. \Cref{fig3}a shows an excerpt from \Cref{figS1}, namely, the marginalization between $\Eap$ and $\sigma_{p0}$. The line of similar posterior probability seen in \Cref{fig3}a (note that $\sigma_{p0}$ is logarithmically spaced) represents the inherent tradeoff between prefactor and activation energy when fitting an exponential model like this over a finite temperature range. This tradeoff is clear from \Cref{fig3}b, which shows the fitted and literature-sourced $\sigma_p$ values at separate temperatures (reproduced from \Cref{fig2}) as well as the lines corresponding to the highest-probability Arrhenius parameter sets from this analysis. 

The dotted line in \Cref{fig3}b represents the Arrhenius fit from Ref.~\citenum{istratov}. However, that fit allowed only the prefactor to vary, fixing the activation energy according to the results of a separate measurement, while in our analysis we allowed the activation energy to be a fitting parameter. A strength of the Bayesian approach is that information from such a measurement can be explicitly incorporated via conditioning the posterior distribution further. If we allow only fits with activation energies near this value (-0.045 eV), which is to say we condition the probability distribution, then \Cref{fig3}a becomes \Cref{fig3}c, and \ref{fig3}b becomes \ref{fig3}d, with the results agreeing even more closely with the literature fit.

In this work, we demonstrate the ability to extract SRH recombination parameters from device-level measurements that yield comparable results to TIDLS and DLTS. In particular, our results fall well within the range of values reported by different DLTS practitioners, and simulated IDLS data are also in agreement. However, our approach utilizes a much simpler and more widely applicable experimental setup -- a temperature-controlled $JV$ stage with a solar simulator and neutral-density filters. Furthermore, $JV$ measurement is a standard industrial characterization technique, meaning this approach could in principle be integrated into production lines. It also shifts a significant number of person-hours of effort to computational resources, which are becoming increasingly inexpensive, plentiful, and user-friendly. In addition, the Bayesian framework allows easy incorporation of any preexisting information from other sources, such as (in this work) parametrization of thermal velocity or prior characterization of trap density or capture barrier. We note that within the range of experimental conditions of our measurements (in particular, all measurements being in the low-injection regime), we were not able to significantly constrain the trap level. This would likely be resolved with a setup capable of concentrated measurements significantly above 1 Sun of illumination.

We emphasize that in any analysis of this kind, the quality of the results obtained is strictly bounded above by the applicability of the model whose parameters are being estimated. For example, if in reality the temperature dependence of capture cross sections deviates from a thermally activated model (as the ``flattening out" of the curves in Figure~\ref{fig2} at low temperatures seems to suggest, the meaning of the associated parameters and their probability distributions could be called into doubt as well.

This work represents a simple, rapid ($\mathcal{O}$(1 day) each experiment time and simulation time on a sufficient HPC cluster) approach to access SRH parameters, which promises to be useful both in screening of novel PV materials as well as characterizing better-known ones, as defect parameter data is generally very sparse in literature due to the complexity of its collection.

\section{Methods}

    \subsection{Experimental Methods}

    For this study, silicon solar cells were obtained from the same set used in previous work where samples were intentionally contaminated with specific amounts of interstitial iron (Fe$_\text{i}$); see Ref.~\citenum{pdg_si} (``60A" samples) for details regarding sample fabrication and measurements of iron concentration. Two of these samples were further characterized in this work: one intentionally contaminated sample with a known Fei concentration of $2\times10^{12}\text{ cm}^{-3}$ (after gettering), and a control sample with no intentional contamination (with estimated [Fe$_\text{i}$] $\le 10^{10}\text{ cm}^{-3}$, based on measurement detection limits). Measurements were first performed on a 1-Sun solar simulator setup (Newport Oriel Sol3A, class AAA, 450 W Xe lamp, AM1.5G filter, Keithley 2400) to verify open-circuit voltage degradation of less than 1.5\% rel. since the samples were first fabricated. Samples were apertured during all $JV$ and $JVTi$ measurements to ensure accurate short-circuit current values would be obtained. Quantum efficiency (PV Measurements QEX7, 300--1100 nm, 75 W Xe lamp, Spectral Products CM110 monochromator) and reflectance data (Perkin-Elmer Lambda 950 UV-Vis spectrophotometer, 150 mm integrating sphere) were also obtained for the purposes of fitting to the PC1D model (see below).

    $JVTi$ measurements were performed under vacuum (approx. $10^{-3}$--$10^{-5}$ Torr) using a liquid helium cryostat (ARS DE-204SI) and compressor (ARS-4HW) to reach colder sample temperatures while avoiding the condensation of atmospheric species; measurements were taken from 300 to 175 K at increments of 25 K. Precise temperature control within $\pm 1$ K was achieved by placing a thermocouple (Omega CY670) directly on the sample surface and using a polyimide resistive heater (Minco HAP6943) and PID temperature controller (Lakeshore 331) to control total heat flux to the sample. Sample illumination at four different intensities (1.01, 0.69, 0.31, and 0.09 Suns, measured with a silicon photodiode) was achieved using a Newport Oriel Solar Simulator (LCS-100, class ABB, 1.5”$\times$1.5” uniform output) along with an array of neutral-density filters placed within two filter wheels (Thorlabs FW102C). $JV$ sweeps were performed using a Keithley 2400 sourcemeter. To ensure all iron present was in the form of Fe$_\text{i}$ (vs. Fe-B pairs), samples were soaked for 15 min at 1 Sun and 300 K before measurements began, as suggested from calculations of temperature-dependent re-pairing rates based on Refs.~\citenum{MacDonald2006} and~\citenum{Zoth1990}.

    \subsection{Computational Methods}
    The 1-Sun $JV$, quantum efficiency, and reflectance measurements were used to construct a numerical device model accessed by the Bayesian inference framework (see below). The use of a modified, command-line version of PC1D~\cite{Haug2015, Haug2016} enabled scripted methods for modifying simulation parameters. Specific input parameters were obtained from previous measurements~\cite{pdg_si}, estimated from literature values, or varied in the model to match the $JV$, QE, and reflectance data of the uncontaminated sample. A full list of device parameters is listed in the Supplementary Information (\Cref{tableS1,tableS2,tableS3}).

    Bayesian inference calculations were performed using the Bayesim package.\cite{bayesim} PC1D simulations were run on MIT Supercloud~\cite{supercloud} using Wine~\cite{wine} and the LLMapReduce~\cite{LLMapReduce} function. Code to reproduce figures plotted herein is available at \url{https://github.com/PV-Lab/Fe_Si_Bayes_code}.

\begin{acknowledgement}

This work was supported by the Center for Next Generation Materials by Design (CNGMD), an Energy Frontier Research Center funded by the U.S. Department of Energy, Office of Science, Basic Energy Sciences, as well as the MIT-Spain - Universidad Politécnica de Madrid Seed Fund and the MIT SuperCloud and Lincoln Laboratory Supercomputing Center for providing HPC and consultation resources that have contributed to the research results reported within this paper. Authors from Aalto additionally acknowledge the provision of facilities and technical support by Aalto University at OtaNano – Micronova Nanofabrication Centre, and the Academy of Finland Flagship Programme, Photonics Research and Innovation (PREIN). J. R. Poindexter acknowledges the support of a Switzer Environmental Fellowship.

We thank Andrei Istratov for helpful discussions, and Lauren Milechin for assistance integrating Wine with LLMapReduce on the Supercloud system.

\end{acknowledgement}

\begin{suppinfo}

\begin{figure}
    \centering
    \includegraphics[width=0.95\textwidth]{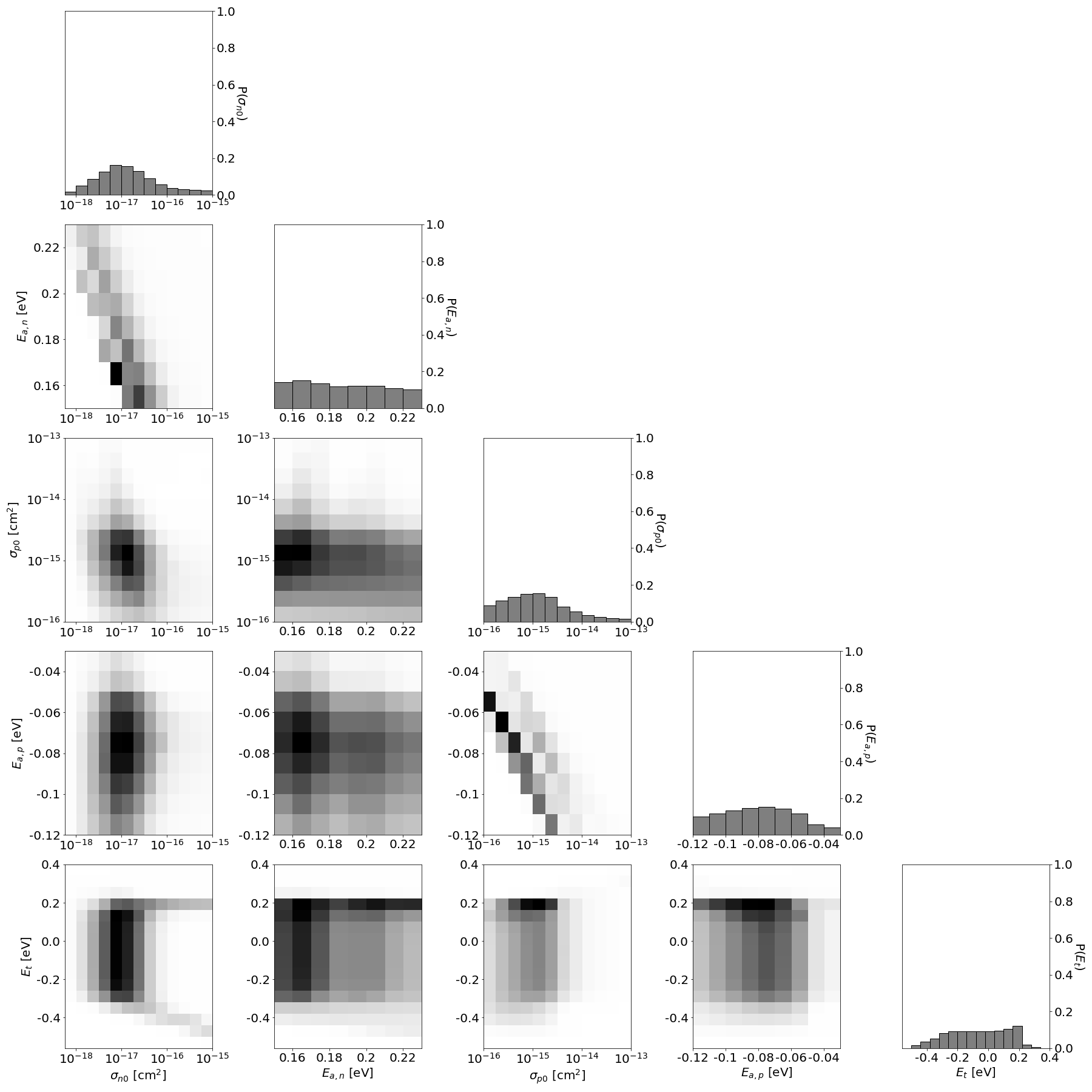}
    \caption{Full five-parameter probability distribution.}
    \label{figS1}
\end{figure}

\begin{table} [!h] \footnotesize
	\begin{center}
		\begin{tabular}{ | p{0.11\textwidth} | p{0.35\textwidth} | p{0.09\textwidth} | p{0.33\textwidth}| } 
			\hline
			\textbf{parameter\newline name} & \textbf{value / setting} & \textbf{Ref.} & \textbf{notes} \\
			\hline
			\hline
			Device Area & 3.55 cm$^2$ & measured 4.00 cm$^2$; Ref. \citenum{pdg_si} & adjusted downward from 4.00 cm$^2$ due to boundary effects (i.e., 1-Sun $J_\text{SC}$ of real cell does not match QE-calculated $J_\text{SC}$). Partially because aperture is used during measurement.) \\ \hline
			
			Surface \newline texture & No surface texturing & & \\ \hline
			
			Surface charge & No surface charge & \citenum{Fell2015} & Ref. \citenum{Fell2015} lists no surface charge for the conventional cell. A rear surface charge of 10$^{10}$ cm$^{-2}$ didn't seem to matter much, either. Also see PC1Dmod 6-2 manual, p. 13. \\ \hline
			
			Reflectance: Front\newline External & Coated; broadband reflectance = 0.69\%; inner layer (thickness, index) = (76 nm, 1.98) & \citenum{pdg_si} & Used high end of thickness, 73$\pm$3 nm, to better fit reflectance data measured experimentally\\ \hline
			
			Reflectance: Rear \newline External & Fixed (0\%) & & \\ \hline
			
			Reflectance: Internal Reflectance & Front surface: specular, 30\% (first bounce and subsequent bounces); Rear surface: specular, 95\% (first bounce and subsequent bounces); & & Adjusted to fit experimentally measured reflectance \\ \hline
			
			Contact \newline definition &
			\footnotesize{
				\begin{tabular} { |p{0.06\textwidth}|p{0.08\textwidth}|p{0.08\textwidth}| }
					\hline
					& internal series resistance & distance from surface \\ \hline
					emitter & 10$^{-8} \Omega$ & 0 $\mu$m \\  \hline
					base & 0.18 $\Omega$ & 400 $\mu$m \\				 
					\hline
				\end{tabular}
			}
			& base and emitter thickness: Ref. \citenum{pdg_si} & \\ \hline
			
			Internal shunt \newline element 1 & conductor, 5.83$\times$10$^{-4}$,\newline anode/cathode/ideality = 400/0/1 & & Fitted to experimental \textit{J--V} data \\ \hline
			
			Global band \newline structure & electron affinity: 4.05 eV & & Other parameters defined by configuration file. \\

			\hline
		\end{tabular}
		\caption{PC1D \textbf{device} parameters for simulating \textit{JVTi} data.}
		\label{tableS1}
	\end{center}
\end{table}
\clearpage

\begin{table} [!h] \footnotesize
	\begin{center}
		\begin{tabular}{ | p{0.25\textwidth} | p{0.3\textwidth} |  p{0.09\textwidth} | p{0.2\textwidth}| } 
			\hline
			\textbf{parameter name} & \textbf{value / setting} & \textbf{Ref.} & \textbf{notes} \\
			\hline
			\hline
			Thickness & 400 $\mu$m & \citenum{pdg_si} & \\ \hline
			
			Dielectric constant & 11.7 & & \\ \hline
			
			Optical properties: \newline Refractive index--External & data file & \citenum{Green2008} & \\ \hline
			
			Optical properties: \newline Intrinsic absorption -- \newline External absorption coeff. & data file & \citenum{Green2008} & \\ \hline
			
			Optical properties: \newline Free-carrier absorption	& Enabled; $\alpha = 2.85\times10^{-26} n\lambda^{2.6} + 1.64\times10^{-25} p\lambda^{2.4}$ & \citenum{Rudiger2013} & \\ \hline
			
			Background doping & \textit{p}-type; 4.979$\times10^{15}$ cm$^{-3}$; resistivity = 2.85 $\Omega$-cm & \citenum{pdg_si} & \\ \hline
			
			First front diffusion & Enabled, \textit{n}-type; calculated from Erfc, sheet resistance = 27.01, junction depth = 1.3 $\mu$m (peak doping / depth factor / and peak position = 1.062e20, 0.4516, 0) & & Calculated in-program. Iterated the sheet resistance and depth factor to match the experimental QE. \\ \hline
			
			Second front diffusion & No second front diffusion & & \\ \hline 
			
			First/second rear diffusion &	No rear diffusion & & \\ \hline
			
			Bulk recombination & fitting parameter & & \\ \hline
			
			Front surface & 1$\times10^7$ cm$/$s, $E_t$ = $E_i$ & suggested from Ref. \citenum{Fell2015} & \\ \hline
			
			Rear surface & 1$\times10^7$ cm$/$s, $E_t$ = $E_i$ & suggested from Ref. \citenum{Fell2015} & \\ \hline
		\end{tabular}
		\caption{PC1D \textbf{material} parameters for simulating \textit{JVTi} data.}
		\label{tableS2}
	\end{center}
\end{table}

\begin{table} [!h] \footnotesize
	\begin{center}
		\begin{tabular}{ |p{0.2\textwidth} |p{0.40\textwidth}| p{0.25\textwidth}| } 
			\hline
			\textbf{parameter name} & \textbf{value / setting} & \textbf{notes} \\
			\hline
			\hline
			Excitation mode & Transient, number of time steps = 100; time step size = 1 s; time step at t=0 = 1e-09 & \\ \hline
			
			Temperature & \textit{input parameter} & \\ \hline
			
			Base circuit & Source: 0 $\Omega$-cm$^2$ resistance; sweep from $-$0.5 to +1.0 V & zero resistance necessary for voltage to sweep full range (vs. some subset) \\ \hline
			
			Collector circuit & all parameters set to zero & \\ \hline
			
			Primary illumination -- intensity & Enable; Front; level is input parameter; AM1.5G spectrum & \\ \hline
			
			Secondary\newline  illumination & disabled & \\
			\hline
		\end{tabular}
		\caption{PC1D \textbf{excitation} parameters for simulating \textit{JVTi} data.}
		\label{tableS3}

	\end{center}
\end{table}

\begin{figure}
    \centering
    \includegraphics[width=0.95\textwidth]{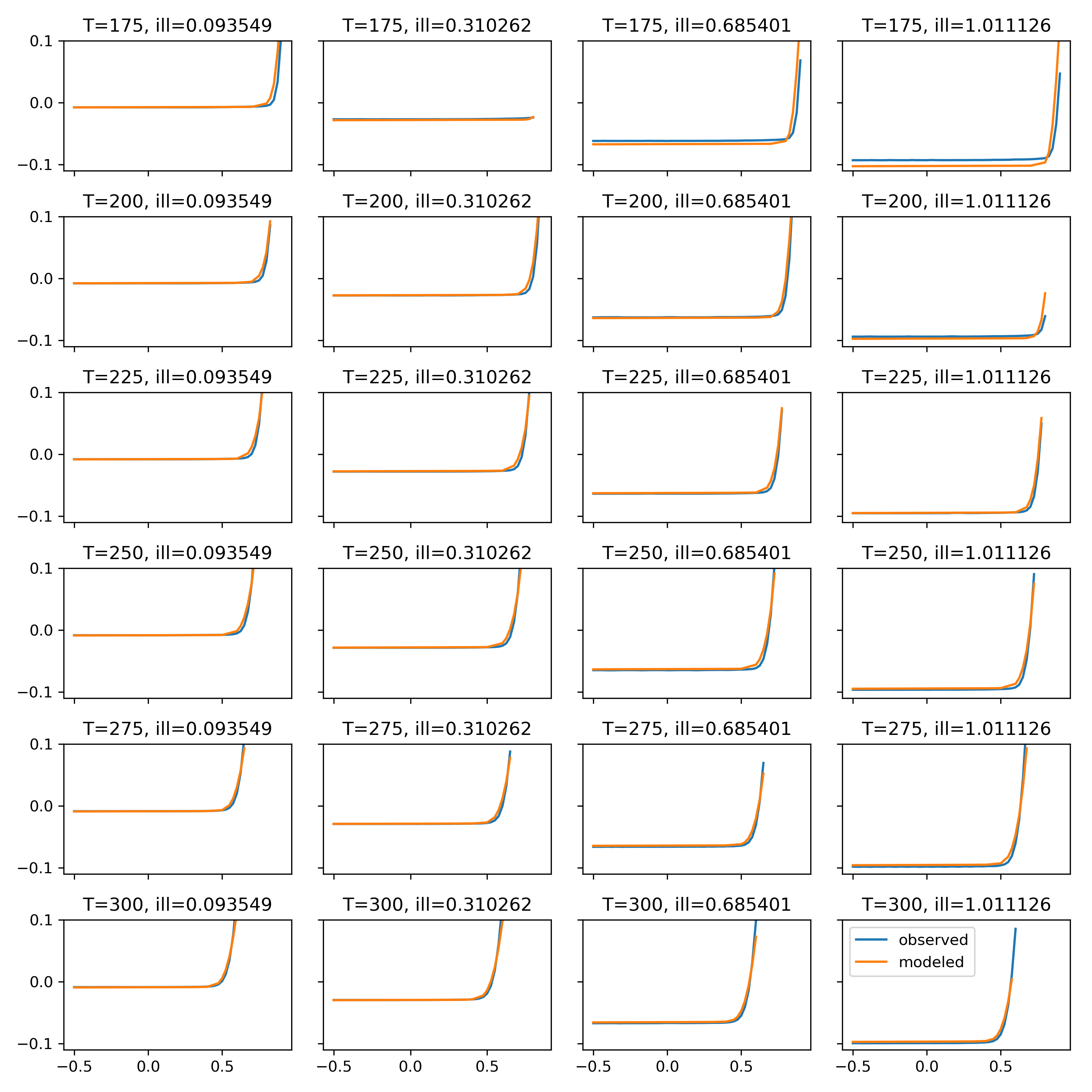}
    \caption{Comparison between modeled and simulated (for highest-probability set of Arrhenius parameters) at every experimental condition. Lack of high-voltage data for some conditions was due to numerical convergence errors.}
    \label{figS2}
\end{figure}

\begin{figure}
    \centering
    \includegraphics[width=0.95\textwidth]{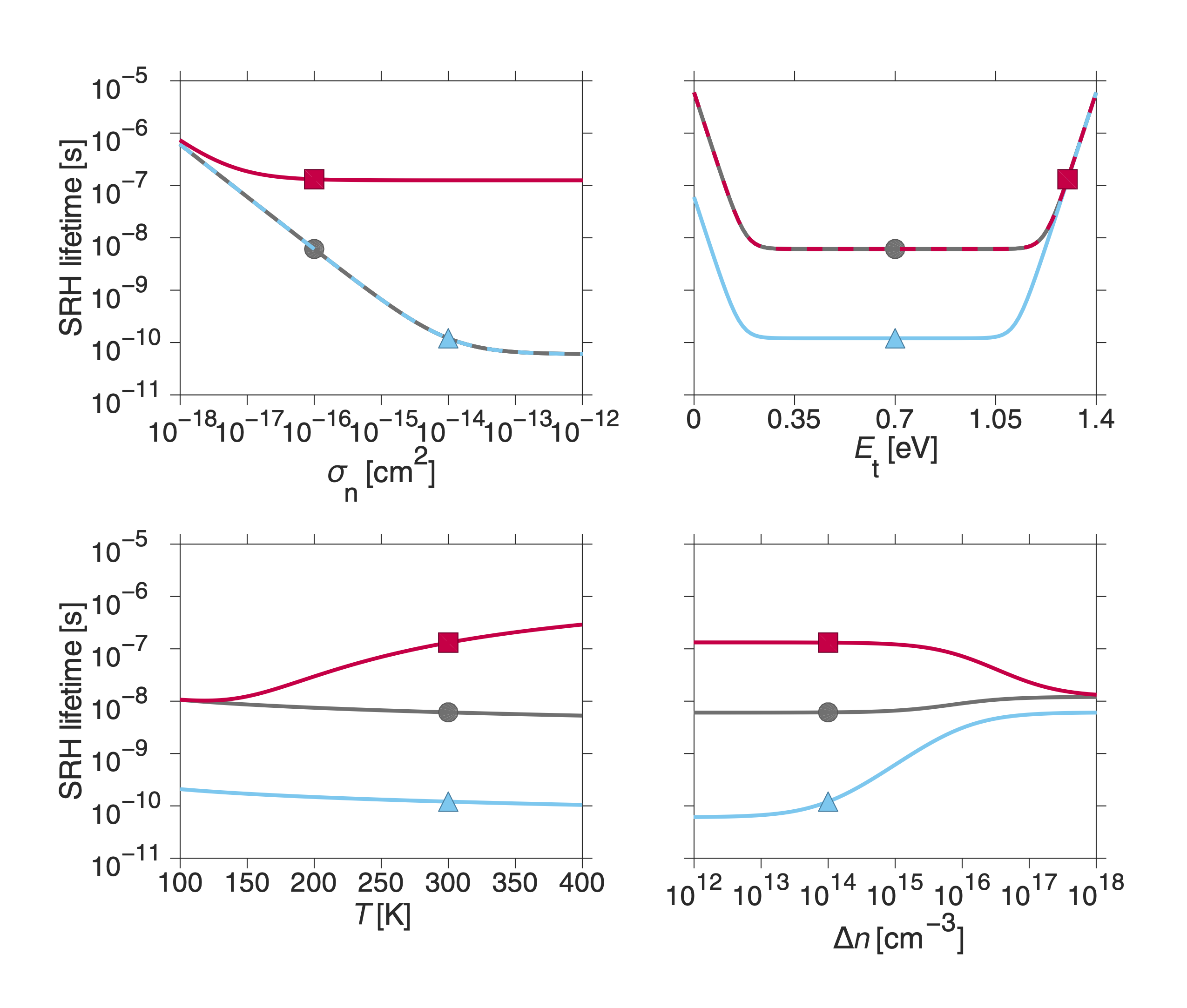}
    \caption{SRH lifetime sensitivity plots showing a baseline calculation (grey dot) along with variations in $\sigma_n$ (blue) and $E_t$ (red), also showing dependence on illumination (injection level $\Delta n$) and temperature.}
    \label{figS3}
\end{figure}

\end{suppinfo}

\bibliography{refs.bib}

\end{document}